**Resolution of the discrepancy between the variation of the physical properties of Ce$_{1-x}$Yb$_x$CoIn$_5$ single crystals and thin films with Yb composition**

S. Jang,[1,2] B. D. White,[2,3] I. K. Lum,[1,2] H. Kim,[4,5] M. A. Tanatar,[4,5] W. E. Straszheim,[4] R. Prozorov,[4,5] T. Keiber,[6] F. Bridges,[6] L. Shu,[7] R. E. Baumbach,[8] M. Janoschek,[9] and M. B. Maple[1,2,3]

[1]Materials Science and Engineering Program, University of California, San Diego, La Jolla, CA 92093, USA

[2]Center for Advanced Nanoscience, University of California, San Diego, La Jolla, CA 92093, USA

[3]Department of Physics, University of California, San Diego, La Jolla, CA 92093, USA

[4]Ames Laboratory, Ames, IA 50011, USA

[5]Department of Physics and Astronomy, Iowa State University, Ames, IA 50011, USA

[6]Physics Department, University of California, Santa Cruz, California 95064, USA

[7]Department of Physics, Fudan University, Shanghai 200433, People's Republic of China

[8]National High Magnetic field Laboratory, Florida State University, Tallahassee, FL 32310, USA

[9]Condensed Matter and Magnet Science, Los Alamos National Laboratory, Los Alamos, NM 87545, USA

July 24, 2014

**Abstract**

The extraordinary electronic phenomena found in the Ce$_{1-x}$Yb$_x$CoIn$_5$ system have attracted much interest. These phenomena include an Yb valence transition, a change in Fermi surface topology, and suppression of the heavy fermion quantum critical field at a nominal concentration $x \approx 0.2$, which, surprisingly, have no discernable effect on the unconventional superconductivity and normal-state non-Fermi liquid behavior that occur over a broad range of $x$ up to ~0.8. However, the variation of the coherence temperature $T^*$ and the superconducting critical temperature $T_c$ with nominal Yb concentration $x$ for bulk single crystals is much weaker than that of thin films. To determine whether differences in the actual Yb concentration of bulk single crystals and thin film samples might be responsible for these discrepancies, we employed Vegard's law and the spectroscopically-determined values of the valences of Ce and Yb as a function of $x$ in Ce$_{1-x}$Yb$_x$CoIn$_5$ flux-grown single crystals to determine the actual composition $x_{act}$ from the variation of the tetragonal a- and c-lattice parameters with $x$. The analysis is supported by energy dispersive x-ray spectroscopy (EDS), wavelength dispersive x-ray spectroscopy (WDS), and transmission x-ray absorption edge spectroscopy (TXAS) measurements. The actual composition $x_{act}$ is found to be about 1/3 of the nominal concentration $x$ up to $x \sim 0.5$, and resolves the discrepancy between the variation of the physical properties of Ce$_{1-x}$Yb$_x$CoIn$_5$ single crystals and thin films with Yb concentration. Measurements of physical properties show that Yb enters the single crystals systematically and in registry with the nominal Yb concentration $x$ of the starting material dissolved in the molten indium flux. As a result, all of the experiments on $Ce_{1-x}Yb_xCoIn_5$ single crystals that have been performed are still valid but for a subnominal actual Yb concentration $x_{act} \sim x/3$, up to $x \sim 0.5$.

**Introduction**

The unusual effect of Yb substituents compared to that of other lanthanide substituents on the normal and superconducting state properties of the heavy fermion compound $CeCoIn_5$ has attracted much recent interest [Capan10, Shu11, Booth11, White12, Polyakov12, Shimozawa12, Hu13, Dudy13, Singh14, Kim14, Song14]. Measurements of the electrical resistivity $\rho(T)$, magnetic susceptibility $\chi(T)$, specific heat C($T$), and tetragonal a- and c-lattice parameters as a function of $x$, performed on flux-grown bulk single crystal specimens in independent studies by Capan $et$ $al$. [Capan10] and Shu $et$ $al$. [Shu11], yielded results that are in general agreement with one another. In the work of Shu $et$ $al$. [Shu11], the weak variations of the coherence temperature $T^*$ and superconducting critical temperature $T_c$ with Yb concentration $x$ were attributed to stabilization of the correlated electron state in the $Ce_{1-x}Yb_xCoIn_5$ system over a large range of $x$. Based on the strong deviation of the tetragonal a- and c-lattice parameters and unit cell volume V = $a^2c$ as a function of $x$ from Vegard's law (linear variation of a and c-lattice parameters with $x$) [Vegard21], it was suggested [Shu11] that the stability of the correlated electron state in $Ce_{1-x}Yb_xCoIn_5$ could be due to cooperative behavior of the Ce and Yb ions involving their unstable valences that can range from +3 to +4 in the case of Ce and +2 to +3 for Yb. Since the 4$f$-electron states of the Ce and Yb ions are admixed with conduction electron states, they communicate with one another through the conduction electrons. It was reasoned that the Ce and Yb ions could then self-consistently adjust their valences so as to stabilize the heavy electron state over a large range of Yb concentrations $x$. However, spectroscopic measurements (EXAFS, XANES, and ARPES) by Booth $et$ $al$. [Booth11] and Dudy $et$ $al$. [Dudy13] found that the valence of Ce remains close to +3 for $0 \le x \le \sim 1$, whereas the valence of Yb remains close to +2.3 for $\sim 0.2 \le x \le 1$. These results are not consistent with the aforementioned proposal [Shu11] that the Ce and Yb valences vary with Yb concentration. It is noteworthy that the experiments of Dudy $et$ $al$. [Dudy13] revealed that Yb undergoes a valence transition from +3 at $x \approx 0$ to $\sim +2.3$ at $x \approx 0.2$. While there are issues of phase separation at values of $x$ above $\sim 0.8$ [Capan10, Shu11], we emphasize that the valence of the pure $YbCoIn_5$ end-member compound ($x = 1$) was found to be $\sim +2.3$ [Booth11, Dudy13], which is important for the analysis presented herein.

Subsequent investigations of the $Ce_{1-x}Yb_xCoIn_5$ system revealed evidence for other electronic transitions at $x \sim 0.2$, including a reconstruction of the Fermi surface above $x \approx 0.2$, accompanied by a significant reduction in the quasiparticle effective mass [Polyakov12], and suppression of the quantum critical field associated with the correlated heavy fermion state to 0 K at a quantum critical point (QCP) at $x \approx 0.2$ [Hu13, Singh14]. Surprisingly, these transitions have little effect on the unconventional superconductivity and non-Fermi liquid (NFL) behavior. Measurements of $T_c$ vs. $x$ in the range $x = 0$ and $x = 0.7$ show that $T_c$ decreases linearly with $x$ from 2.3 K at $x = 0$ and extrapolates to 0 K at $x \approx 1$, with no features near $x = 0.2$. The NFL signatures in $\rho(T)$, $C(T)$, and $\chi(T)$ persist from 0 to $x \approx 0.8$ with an abrupt crossover to Fermi liquid (FL) behavior slightly above this value of $x$ [Shu11, Singh14]; this suggests that the NFL behavior could be associated with a new state of matter rather than being a consequence of the underlying quantum phase transition at $x \approx 0.2$ and opens up the possibility that some other type of electronic transition occurs near $x \approx 0.8$ [Erten14]. The region $0.8 \leq x \leq 1$ is currently being explored.

Measurements of electrical transport properties on $Ce_{1-x}Yb_xCoIn_5$ thin films revealed that the variation of the physical properties is much stronger than that observed in the $Ce_{1-x}Yb_xCoIn_5$ single crystals, but weaker than that which is found for other lanthanide substituents in $CeCoIn_5$ single crystals [Shimozawa12]; the rate of depression of $T^*$ and $T_c$ with Yb concentration for the thin films is about three times greater than that observed in the single crystals. In this paper, we describe an analysis we have performed on bulk single crystals of $Ce_{1-x}Yb_xCoIn_5$ that apparently resolves the discrepancy between the bulk single crystal and thin film experiments. The analysis involves the application of Vegard's law to estimate the actual composition $x_{act}$ using the variation of the tetragonal a- and c-lattice parameters and the spectroscopically-determined valences of Ce and Yb as a function of the nominal composition $x$ in flux-grown $Ce_{1-x}Yb_xCoIn_5$ single crystals. The analysis is unusual, since we are using the known valences of Ce and Yb as a function of $x$ to determine the actual Yb concentration $x_{act}$ (assuming that all of the lanthanide sites in the compounds are occupied by Ce or Yb ions; i.e., no lanthanide vacancies). In the usual Vegard's law analysis, the lanthanide sites are occupied by lanthanide ions according to their nominal concentrations, and the deviations of the lattice parameters from Vegard's law are used to estimate the valence of one of the lanthanide ions [Maple71, Maple74]. Direct support for the Vegard's law analysis is provided by energy dispersive x-ray spectroscopy (EDS), wavelength dispersive x-ray spectroscopy (WDS), and transmission x-ray absorption edge spectroscopy (TXAS) measurements on selected single crystals, which are reported herein. Although XAS usually refers to both fluorescence and transmission data, only the step height in transmission data is proportional to the number of atoms within the x-ray beam and we use the designation TXAS to highlight this difference. The Vegard's law analysis of the a- and c-lattice parameters indicates that the actual Yb composition of the single crystals is about 1/3 of the nominal composition in the range $0 \leq x \leq {\sim}0.5$, resolving the discrepancy between experiments on $Ce_{1-x}Yb_xCoIn_5$ single crystals and thin films. It is noteworthy that the actual composition of the $Ce_{1-x}Yb_xCoIn_5$ single crystals prepared from a molten indium flux is in registry with the nominal composition of the starting material contained in the molten flux, but is only about 1/3 of its value in the range $0 \leq x \leq {\sim}0.5$. The sharpness of the specific heat feature associated with the superconducting transition [Capan10, Shu11], which reflects the bulk behavior of the crystals, indicates that the $Ce_{1-x}Yb_xCoIn_5$ single crystals are homogeneous up to Yb nominal compositions of ${\sim}0.5$. As a result, all of the experiments on $Ce_{1-x}Yb_xCoIn_5$ single crystals that have been performed in the range $0 \leq x \leq {\sim}0.5$ are still valid but the actual

Yb concentration is about 1/3 of the nominal concentration; i.e., $x_{act} \approx x/3$. At higher $x$, the Vegard's law analysis indicates that the ratio of $x_{act}$ to $x$ increases continuously with $x$ to the value of 1 at $x = 1$, as it must for the pure $YbCoIn_5$ end member compound.

In the study by Shu $et$ $al$. [Shu11], the samples selected for $\rho(T)$, $\chi(T)$ and C($T$) measurements had actual Yb compositions $x_{act}$, as determined from EDS measurements, that were close to the nominal composition $x$. We have performed further EDS measurements on samples with the same values of $x$ as those reported in the work of Shu $et$ $al$. and found that the values of $x_{act}$, inferred from EDS measurements, exhibit bimodal behavior in which $x_{act} \approx x/3$ for some crystals, whereas $x_{act} \approx x$ for other crystals, with large uncertainties in the values of $x_{act}$. A possible reason for this bimodal behavior of $x_{act}$ may be that the single crystals have a bimodal Yb surface composition $x_{act} \approx x/3$ and $x_{act} \approx x$ and a bulk composition $x_{act} \approx x/3$. As we explain in the following, this result is consistent with an analysis based on the application of Vegard's law to the unit cell volume, $V$, vs. $x$ data, where $V$ was determined from powder x-ray diffraction measurements of the tetragonal a- and c-lattice parameters (i.e, $V = a^2c$).

**Experimental details**

Single crystals of $Ce_{1-x}Yb_xCoIn_5$ were grown using a molten indium flux method in alumina crucibles, as described previously [Zapf01]. Powder x-ray diffraction measurements, performed at room temperature, reveal that the $Ce_{1-x}Yb_xCoIn_5$ single crystals form in the tetragonal $HoCoGa_5$ structure. The tetragonal a- and c-lattice parameters were determined from a least-squares fit of the peak positions in the x-ray powder diffraction pattern using GSAS and EXPGUI [Rietveld69, Larson00]. It should be noted that our lattice parameter values are in good agreement with those of Capan $et$ $al$. [Capan10]; however, in the analysis described below, we use the data for the a- and c-lattice parameters of Capan $et$ $al$., since the scatter in their data is smaller than in ours.

Energy dispersive x-ray spectroscopy (EDS) measurements were carried out at the University of California, San Diego on various samples that have been prepared for the experiments reported in Refs. [Shu11, White12, Hu13, Dudy13, Singh14, Kim14, Song14], the wavelength dispersive x-ray spectroscopy (WDS) measurements were performed at Iowa State University, Ames National Laboratory, on samples studied in the penetration depth experiments and on several other samples with different Yb concentrations [Kim14], while the transmission x-ray absorption edge spectroscopy (TXAS) measurements were performed at the Stanford Synchrotron Radiation Laboratory on samples specifically prepared for the TXAS measurements. The EDS, WDS, and TXAS measurements are discussed in more detail below.

Measurements of electrical resistivity were performed down to ~1.1 K in a $^4$He Dewar using a Linear Research LR700 ac resistance bridge. Four wires were adhered to gold-sputtered contact pads on each single crystal using silver epoxy. Typical contact resistances of 100 m$\Omega$ or less were measured at room temperature by comparing two- and four-wire resistance measurements.

**Results and Discussion**

Before determining the actual Yb concentration $x_{act}$ in single-crystalline samples of $Ce_{1-x}Yb_xCoIn_5$, we demonstrate that Yb is incorporated into the single crystals systematically and in registry with the nominal Yb concentration $x$. In Fig. 1(a), the inverse of the residual resistivity ratio (RRR) is plotted vs $x$. We calculated the inverse of the RRR from electrical resistivity $\rho(T)$ measurements (not shown) on several single crystals as $\rho_0/\rho(300\ K)$, where $\rho_0$ is the residual electrical resistivity. The data are plotted this way because $\rho_0/\rho(300\ K)$ does not suffer from errors associated with measuring the geometrical factor. It is clear that the data in Fig. 1(a) are linear up to $x = 0.5$, convincingly demonstrating that Yb is incorporated into the crystal structure systematically (i.e., temperature-independent impurity scattering characterized by $\rho_0$ increases relative to $\rho(300\ K)$ with increasing nominal Yb concentration $x$). We are also able to demonstrate that the variation of the physical properties of the $Ce_{1-x}Yb_xCoIn_5$ system is a systematic function of the RRR as seen in the plot of superconducting critical temperature $T_c$ vs. RRR in Fig. 1(b). This result further emphasizes that the RRR values meaningfully characterize the level of disorder in the single crystals, and that $x_{act}$ is in registry with $x$.

*Estimate of actual Yb concentration using Vegard's law*

Vegard's law refers to the linear variation of the lattice parameters or unit cell volume when an element is substituted for another element in a compound for which the crystal structure does not change and the chemical composition is known. This relation is often used to estimate the valence of lanthanide ions, which have valence instabilities such as Ce, whose valence can range from 3+ to 4+, and Sm, Eu, Tm, and Yb, whose valence can range from 2+ to 3+ [Maple71, Maple74]. In this work, we instead employ Vegard's law in a different manner to estimate the actual Yb concentration $x_{act}$ in the $Ce_{1-x}Yb_xCoIn_5$ system using the valences of the Ce and Yb ions, derived from spectroscopic measurements, and the tetragonal a- and c-lattice parameters, determined from x-ray diffraction measurements on powdered $Ce_{1-x}Yb_xCoIn_5$ samples, as a function of nominal Yb concentration $x$. Since previous investigations have shown that the valence of Yb changes from 3+ to 2.3+ between $x = 0$ and $x = 0.2$ [Dudy13] and remains 2.3+ for all concentrations greater than $x = 0.2$ [Booth11, Dudy13], the unit cell volume $V = a^2c$ of $Ce_{1-x}Yb_xCoIn_5$ should be a linear function of $x$ between 0.2 and 1.0, according to Vegard's law. We would expect a subtle non-linear variation of $V$ with $x$ for $x < 0.2$ as the Yb valence changes from 3+ to 2.3+ in the range $0 < x \leq 0.2$. In Fig. 2, we show a plot of the measured unit cell volume $V$ vs. $x$, based on the x-ray diffraction measurements of a and c-lattice constants for $Ce_{1-x}Yb_xCoIn_5$ as a function of $x$, and a plot of $V$ vs. $x$ based on Vegard's law. The lattice parameter data of Capan *et al*. [Capan10] were used since they show less scatter than the data of Shu *et al*. [Shu11]. By adjusting the measured values of $V$ to the Vegard's law curve (illustrated by the arrows in Fig. 2), we can estimate the actual concentration $x_{act}$ of Yb. The values of $x_{act}$ determined by means of this procedure are shown in the $x_{act}$ vs. $x$ plot in Fig. 3(b). For values of $x$ below

~0.5, the actual bulk Yb concentration $x_{act}$ is about 1/3 of the nominal concentration $x$. For comparison, Fig. 3(a) shows $x_{act}$ vs $x$ data based on EDS and PIXE measurements on flux grown $Ce_{1-x}Yb_xCoIn_5$ single crystals prepared at UC, Irvine and UC, San Diego. In Fig. 3(a), the $x_{act}$ vs $x$ data are consistent with $x_{act} \approx x/3$ for $x$ values up to $x \approx 0.6$. Shown in Fig. 3(b) are measurements of the bulk concentration of Yb based on EDS, WDS, TXAS, described in the following, that are also seen to be consistent with $x_{act} \approx x/3$ up to $x \approx 0.5$.

*EDS, WDS, and TXAS measurements of the actual Yb concentration*

*Energy dispersive x-ray spectroscopy (EDS) measurements*

The results of EDS measurements on selected $Ce_{1-x}Yb_xCoIn_5$ samples are shown in the $x_{act}$ vs. $x$ plot in Fig. 3(b). The method in which the data were taken is illustrated in Fig. 4, which shows photographs of several $Ce_{1-x}Yb_xCoIn_5$ samples with nominal Yb concentrations of $x = 0.175$ and 0.2, labeled with the letters A, B, C, and D, that have been affixed to conducting carbon tape for the EDS measurements. The EDS measurements were made on each of the samples at several different spots defined by the regions outlined by the black rectangles on the photographs of the crystals. The measured values of $x$, $x_{meas}$, for each of the regions are indicated in the lower panel of Fig. 4 for the crystals labeled A, B, C, and D for the two selected Yb compositions, where the dashed lines indicate the average value of all the data for each Yb nominal concentration. The average values of the measurements and the error derived from the standard deviation are plotted as $x_{act}$ vs. $x$ in Fig. 3(b).

*Wavelength dispersive x-ray spectroscopy (WDS) measurements*

Wavelength dispersive x-ray spectroscopy (WDS) measurements were performed using a JEOL JXA-8200 electron microprobe on all samples used in the London penetration depth study [Kim14] and several representative samples at other nominal concentrations of Yb, including samples with Yb from different sources. The composition of each single crystal was measured at twelve different locations on typically 0.5 x 0.5 mm$^2$ samples and averaged, yielding statistical error of compositional measurement of about $\Delta x = \pm 0.005$. This error bar is substantially smaller than that of the EDS measurements because of the weakness of the Yb line, which causes the EDS measurements to have significantly lower spectral resolution and a small signal to noise ratio. The M-lines of Yb as measured in materials with $Yb^{3+}$, $YbF_3$ and $YbRh_2Si_2$ are shown in Fig. 5(a) and in Fig. 5(b) for data normalized at the peak position. The data for $YbCoIn_5$ and $Ce_{1-x}Yb_xCoIn_5$ with a nominal Yb concentration of $x = 0.40$ are also shown in Figs. 5(a) and (b). Two features in the data should be noted. First, the M-lines in compounds with $Yb^{3+}$ are slightly shifted with respect to the lines in both $YbCoIn_5$ and $Ce_{1-x}Yb_xCoIn_5$, where the lines coincide. These observations are consistent with an Yb valence that is different from 3+ in $Ce_{1-x}Yb_xCoIn_5$ compounds for the whole series and are in agreement with previous spectroscopic measurements of the Yb valence [Booth11, Dudy13]. This result provides direct evidence that Yb substitution induces hole doping, similar to Cd substitution [Pham06]. Furthermore, both direct comparison of the spectra in Fig. 5(a) and more quantitative analysis of $x$, taking into account mutual absorption, show that the actual Yb concentration $x_{act}$ is proportional to and smaller in magnitude than the nominal Yb concentration $x$, roughly by a factor of 3, as shown in Fig. 3.

*Transmission X-ray absorption spectroscopy (TXAS) measurements*

To determine the Ce and Yb concentrations relative to Co from x-ray absorption spectroscopy, the absorption step heights at the Ce and Yb $L_{III}$ edges and the Co K edge were measured in transmission.   The advantage of transmission measurements is that they do not just detect the atoms near the surface of the material, but instead probe all the atoms of interest, since the x-ray beam passes through the sample.  The drawback is that they also detect atoms associated with inclusions that consist of the flux (in this case, indium) or impurity phases.  For each thin powdered sample, transmission data were collected at the same point on the sample.  The transmission step height is a direct measure of the number of atoms in the beam; using thin layers minimizes the effects of sample pinholes and inclusions.   We have also checked that using a thicker sample (twice as thick) gives the same step-height ratios.  Examples at the Yb $L_{III}$ and Co K-edge are shown in Fig.  6.  To obtain the atomic ratios, we first normalize each step by the known absorption step per atom for that element, from the MacMaster x-ray absorption cross-sections [MacMaster69]; then the atomic ratios are given by the ratios of these normalized step heights.   Assuming the Co site is fully occupied, this ratio directly gives the Yb or Ce concentration.   The errors in such concentration measurements are about 5%. A similar approach was used to determine the Zn concentration in Zn doped LiNbO$_3$ [Bridges12]. The measured Ce concentrations are close to the nominal values, but the Yb concentrations are much lower than expected. The Yb concentrations are plotted in Fig. 3(b), which shows $x_{act}$ vs. $x$, for two samples with nominal Yb concentrations of 0.3 and 0.4.

*Correction of the T vs. Yb concentration phase diagram*

The values of the actual Yb concentration, $x_{act}$, determined from the Vegard's law procedure and supported by the measurements of $x_{act}$ described above, can now be used to correct the $T$ vs. Yb concentration phase diagram proposed in Refs. [Capan10, Shu11].  In Fig. 7, the values of the superconducting critical temperature $T_c$ are shown vs. the actual concentration $x_{act}$, and compared to the data for the studies on thin films by Shimozawa *et al*. [Shimozawa12].  The $T_c$ vs. $x_{act}$ phase boundary for both the bulk single crystal and thin film measurements are in good agreement with one another, which supports the procedure to determine the actual Yb concentration we have employed in this work.  A revised phase diagram that includes the $T_c$ vs. $x_{act}$ phase boundary and the dependence of the Kondo lattice coherence temperature $T^*$ on $x_{act}$ is displayed in Fig. 8. The only remaining discrepancy between the behavior of bulk single crystals and thin film samples is that, while $T^*$ is monotonically suppressed in the thin film samples with increasing $x_{act}$, $T^*$ increases with increasing $x_{act}$ above $x_{act} \sim 0.25$ in single crystals.  While we do not understand the origin of this discrepancy, we note that the thin film samples differ from the single crystals through the presence of the substrate, arrested a-lattice parameter, etc. [Shimozawa12].

This revision of the phase diagram of $Ce_{1-x}Yb_xCoIn_5$ does not change any of the interesting physics that has been found for this extraordinary system, but simply readjusts the concentration at which various phenomena occur. In particular, we now conclude that the valence transition of Yb from 3.0+ to 2.3+ occurs between $x = 0$ and ~0.07.

**Concluding remarks**

The analysis involving Vegard's law and the spectroscopically-determined valences of Ce and Yb as a function of the nominal concentration $x$ in the $Ce_{1-x}Yb_xCoIn_5$ system have yielded estimates of the actual Yb concentration $x_{act} \approx x/3$ for x below ~0.5. The relation between $x_{act}$ and $x$ derived from this analysis is supported by the results of EDS, WDS and TXAS measurements reported in this work.

The subnominal Yb concentration encountered in the flux grown $Ce_{1-x}Yb_xCoIn_5$ single crystals is not without precedent for other substituents in $CeCoIn_5$. For instance, Cd, Hg, and Sn substitution on the indium site occurs at fractional values of the nominal concentration [Pham06, Bauer05, Gofryk12]. In addition, there is a strong preference for substitution on the In(1) site [Booth09], which has implications for how the electronic structure is tuned. Similar obstacles are also seen for Pt and Ru substitution [Gofryk12, Ou13]. Moreover, in the case of $CeCo_{1-x}Ru_xIn_5$, not only is the measured $x$ less than the target $x$, but it is also seen that the Ru ions form clusters in the 115 lattice (at least on the surface of the flux grown crystals) [Ou13]. There is evidence that such problems may also be found in other transition metal substitution series: e.g., the broad region of coexistence of antiferromagnetism and superconductivity in the *T-x* phase diagram for $CeCo_{1-x}Rh_xIn_5$ may suggest phase separation in localized regions of individual crystals [Jeffries05]. An exception to these difficulties is seen for lanthanide substitution of the Ce site (e.g., in $Ce_{1-x}R_xCoIn_5$: R = Y, La, Pr, Nd, Gd, Dy, Er, and Lu), which appears to be straightforward [Paglione07, Hu08, Petrovic02].

However, as we have described above, R = Yb substitution is problematic and occurs only at fractional values of the nominal concentration, similar to what is seen for substitution on the indium and transition metal sites. This result accounts for our earlier reports of an exceptional *T-x* phase diagram for $Ce_{1-x}Yb_xCoIn_5$ and, upon rescaling of $x$, our phase diagram is now in close agreement with that of thin films of $Ce_{1-x}Yb_xCoIn_5$. Notably, it is important and interesting that the actual concentration $x_{act}$ is in registry with the nominal concentration $x$ of the starting material in the molten In flux. As a result, none of the interesting physics that has been found for this extraordinary system is changed; the various electronic transitions and phenomena simply occur at an actual Yb concentration $x_{act} \approx x/3$ for below ~0.5. In particular, we now conclude that the valence transition of Yb from 3.0+ to 2.3+ occurs between $x = 0$ and ~0.07. As we noted above, Vegard's law is ordinarily used to determine changes of valence of lanthanide ions with unstable valence when the lanthanide sites are completely occupied by lanthanide ions. In the present case, we use the known valences of the Ce and Yb ions as a function of $x$, determined from spectroscopy, and Vegard's law to estimate the actual concentration $x_{act}$ of the Yb ions. This resolves the discrepancy between the variation of the coherence temperature $T^*$ and the superconducting critical temperature $T_c$ with Yb substituent concentration in the $Ce_{1-x}Yb_xCoIn_5$ bulk single crystals and thin films.

**Acknowledgements**

Research at UCSD was supported by the U.S. Department of Energy under Grant No. DE-FG02-04ER46105. Work at Ames Laboratory was supported by the U.S. Department of Energy (DOE), Office of Science, Basic Energy Sciences, Materials Science and Engineering Division. Ames Laboratory is operated for the U.S. DOE by Iowa State University under Contract No. DE-AC02-07CH11358. Research at UCSC was supported by the National Science Foundation under Grant No. DMR1005568. The experiments were performed at SSRL, operated by the DOE, Division of Chemical Sciences. M. J. acknowledges support by the Alexander von Humboldt Foundation. I. K. L. was supported by a Quantum Design Ronald E. Sager Fellowship. It is a pleasure to thank Prof. Carmen Almasan, Prof. James Hamlin, and Dr. Kevin Huang for useful discussions.

**Figure Captions**

Figure 1: (a) Inverse of the residual resistivity ratio (RRR), calculated as $\rho_0/\rho(300 \text{ K})$ from measurements of electrical resistivity on distinct samples, plotted vs. nominal Yb concentration $x$ for the $Ce_{1-x}Yb_xCoIn_5$ system. The dashed line is a guide to the eye. (b) Superconducting critical temperature $T_c$ plotted vs. RRR from the same measurements shown in panel (a). The dashed curve is a guide to the eye. Vertical bars characterize the width of the superconducting transitions and were calculated using the temperatures where $\rho(T)$ drops to 90% and 10% of its normal-state value just above $T_c$.

Figure 2: Unit cell volume $V$ vs. nominal Yb concentration $x$, for the system $Ce_{1-x}Yb_xCoIn_5$ (filled circles) based on the data of Capan *et al.* [Capan10]. The dashed straight line represents Vegard's law between the unit cell volumes of the end member compounds $CeCoIn_5$, in which the Ce ion has a valence of 3+, and $YbCoIn_5$, in which the Yb ion has an intermediate valence of 2.3+. Since the Yb ion has a valence of 2.3+ in the range $0.2 \leq x \leq 1$, the unit cell volume $V$ should conform to the dashed line in this range of $x$ values, with small deviations in the range $0 < x < 0.2$ where the Yb ions undergo a valence transition between 3+ and 2.3+. The actual concentration of Yb, $x_{act}$, can be estimated by displacing the observed $V(x)$ data (solid circles) to the left, as illustrated by the horizontal arrows, so that they lie on the linear Vegard's law relation (dashed black line).

Figure 3: (a) Actual Yb concentration $x_{act}$ vs. nominal Yb concentration $x$ as obtained from EDS measurements reported by Dudy *et al.* [Dudy13] (solid black squares) and PIXE and EDS measurements reported by Capan *et al.* [Capan10] (solid red circles and solid

inverted blue triangles, respectively). (b) Actual Yb concentration $x_{act}$ vs. nominal Yb concentration $x$ based on: (1) Vegard's law analysis of $V(x)$ measurements (solid red triangles); (2) average value of the measured Yb concentration from several EDS measurements with error bars defined as the standard deviation (solid black circles); (3) WDS measurements (inverted unfilled triangles); (4) TXAS measurements (solid blue squares).

Figure 4. Illustration of method for acquiring EDS data on the $Ce_{1-x}Yb_xCoIn_5$ single crystals. Two samples with $x = 0.175$ and $0.2$ were affixed to conducting carbon tape for EDS measurements labeled with the letters A, B, C, and D. The EDS measurements were made on each sample within several different regions with varying sizes enclosed by the black rectangles. The results of the measurements are plotted as the red and black solid circles for $x = 0.175$ and $0.2$, respectively. The dashed lines represent the average value of the data and are denoted as $x_{act}$ in Fig. 3 for each of the two concentrations.

Figure 5: Results of wavelength dispersive x-ray spectroscopy (WDS) measurements performed on samples used in a penetration depth study [Kim14] for two representative samples with nominal Yb concentrations $x = 0.4$ and $x = 1$ and two compounds in which Yb is trivalent, $YbF_3$ and $YbRh_2Si_2$. (a) M-line of Yb for the four compounds. (b) Data in Fig. 5 (a) normalized at the peak position. Note that the M-lines in the compounds with $Yb^{3+}$ are slightly shifted with respect to the lines in both $YbCoIn_5$ and $Ce_{1-x}Yb_xCoIn_5$, where the lines coincide. These observations are consistent with an Yb intermediate valence of 2.3+ for $Ce_{1-x}Yb_xCoIn_5$ in the range $0.2 \leq x \leq 1$.

Figure 6: Plots of the Co K (a) and Yb $L_{III}$ (b) edges after a linear pre-edge subtraction, which sets the pre-edge region at zero. A straight line fit above the edge, shown as a black line, provides an estimate of the step height at the edge. The data just above each edge that contain the XANES structure were not included in this straight line fit.

Figure 7: Superconducting critical temperature $T_c$, determined from measurements of $\rho(T)$, vs $x$ for $Ce_{1-x}Yb_xCoIn_5$. The data of Shu $et\ al.$ [Shu11] (filled circles) and Capan $et\ al.$ [Capan10] (filled triangles) for flux grown single crystals are plotted vs. $x$, where $x$ represents the nominal concentration (upper horizontal axis). The unfilled circles and triangles indicate those same $T_c$ values plotted vs. $x_{act}$, where $x_{act} \approx x/3$ represents the actual Yb concentration (lower horizontal axis). The solid line indicates the evolution of $T_c$ with $x$ for thin film samples reported by Shimozawa $et\ al.$ [Shimozawa12]. In the case of the thin films, the actual concentration is identical to the nominal concentration $x$.

Figure 8: Coherence temperature $T^*$ and superconducting critical temperature $T_c$ vs. actual Yb composition, $x_{act}$, for flux grown single crystals [Shu11] and thin films [Shimozawa12]. The actual Yb concentration $x_{act}$ for the single crystals was inferred from the nominal concentration as described in the text, while $x_{act}$ for the thin films is the same as the nominal Yb composition.

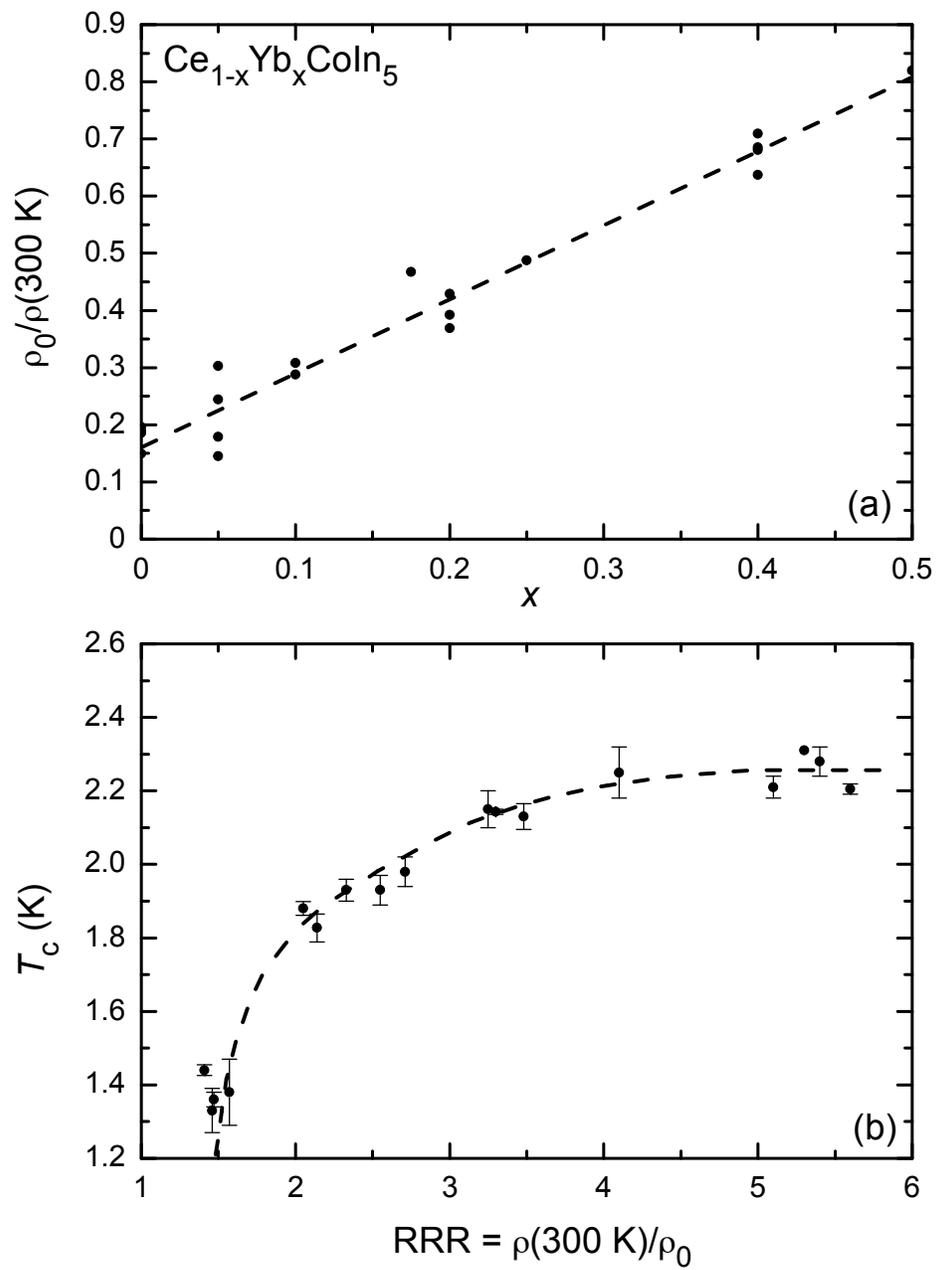

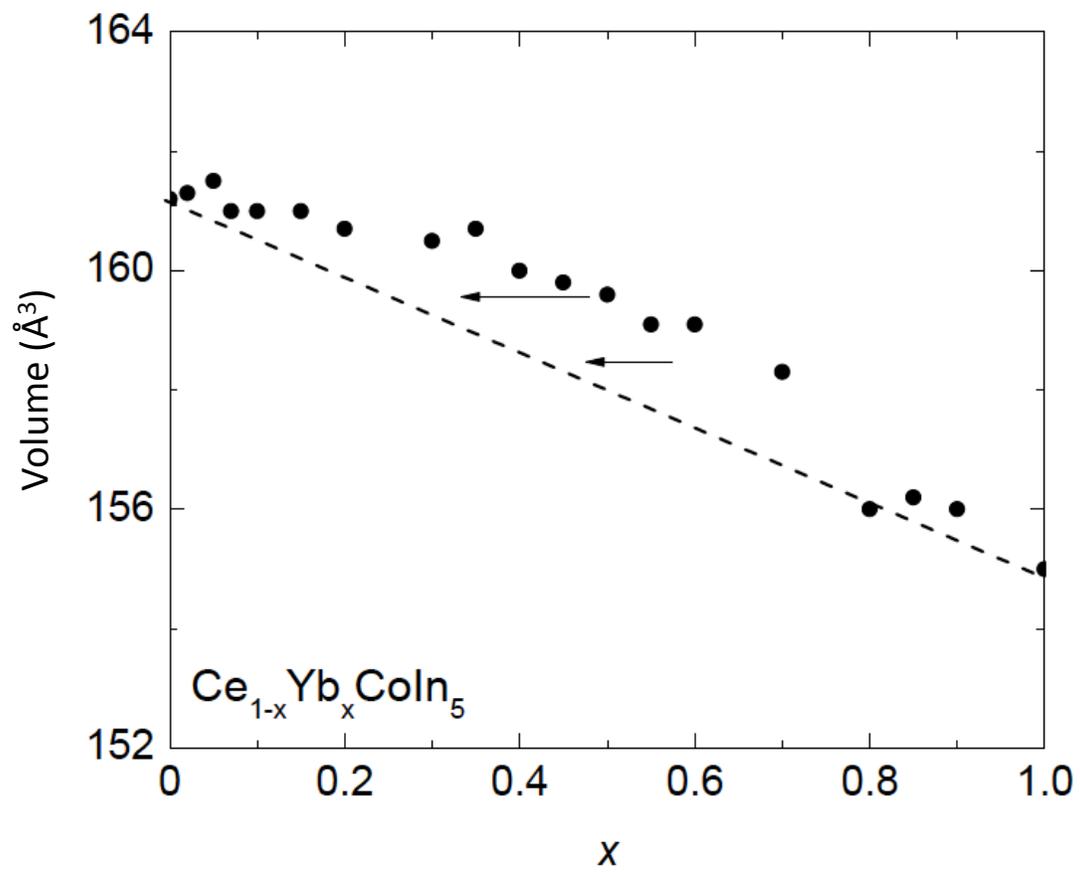

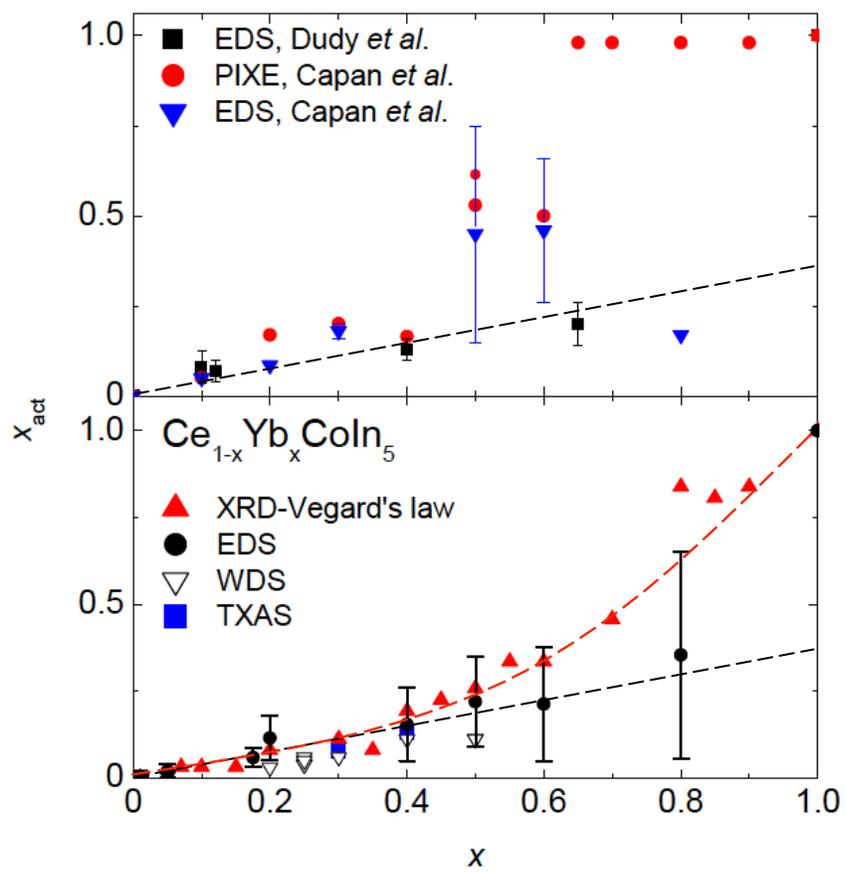

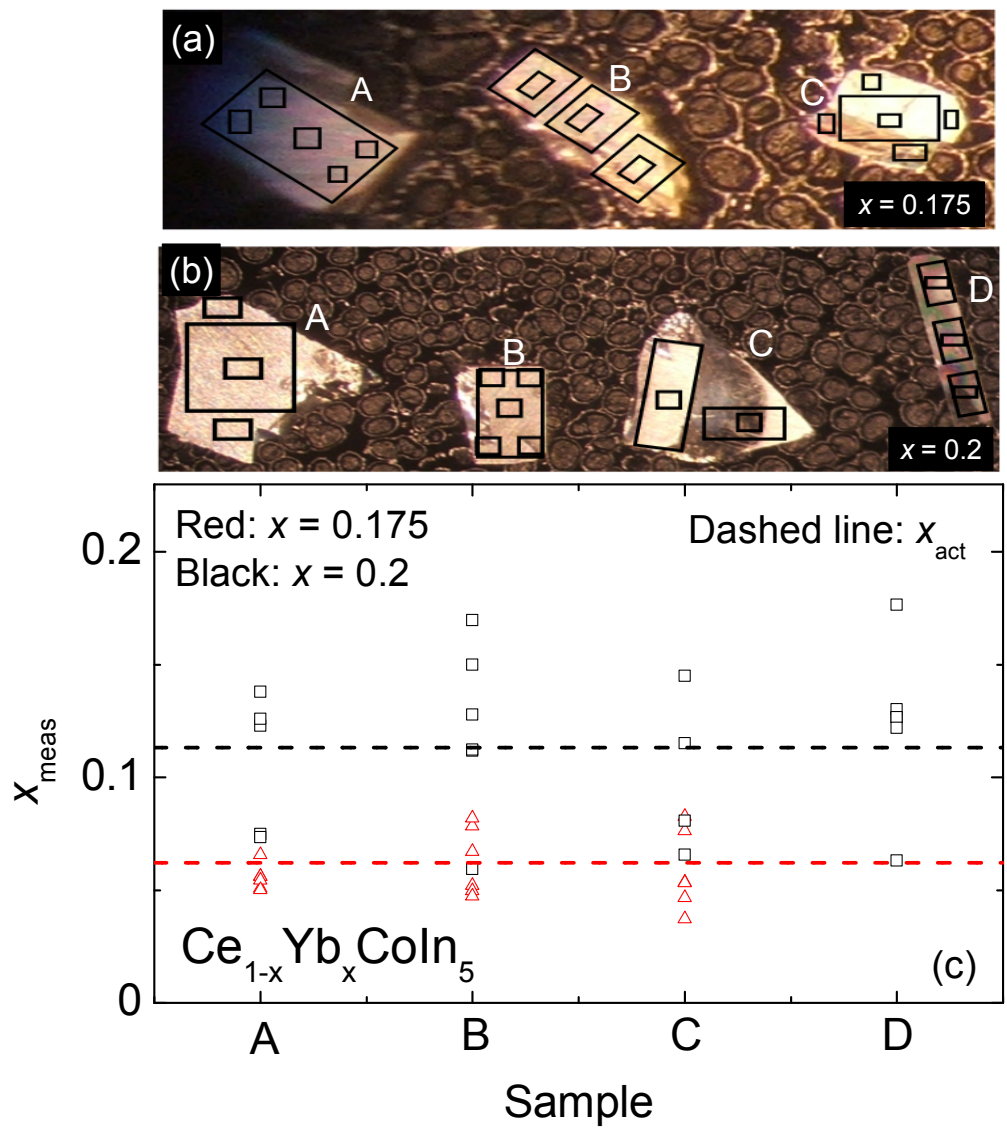

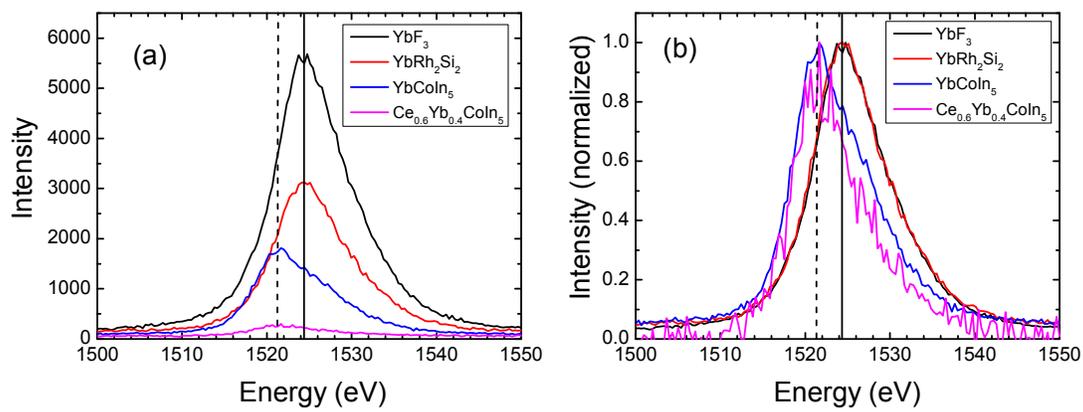

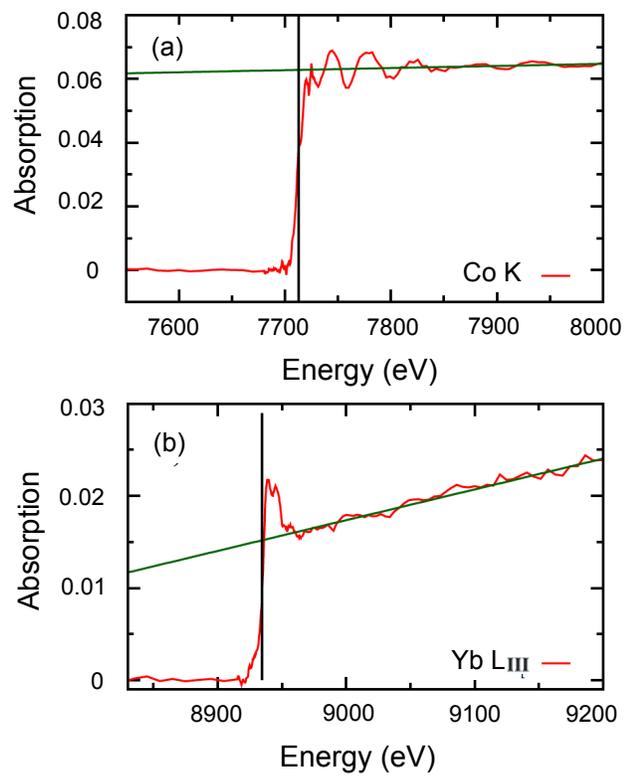

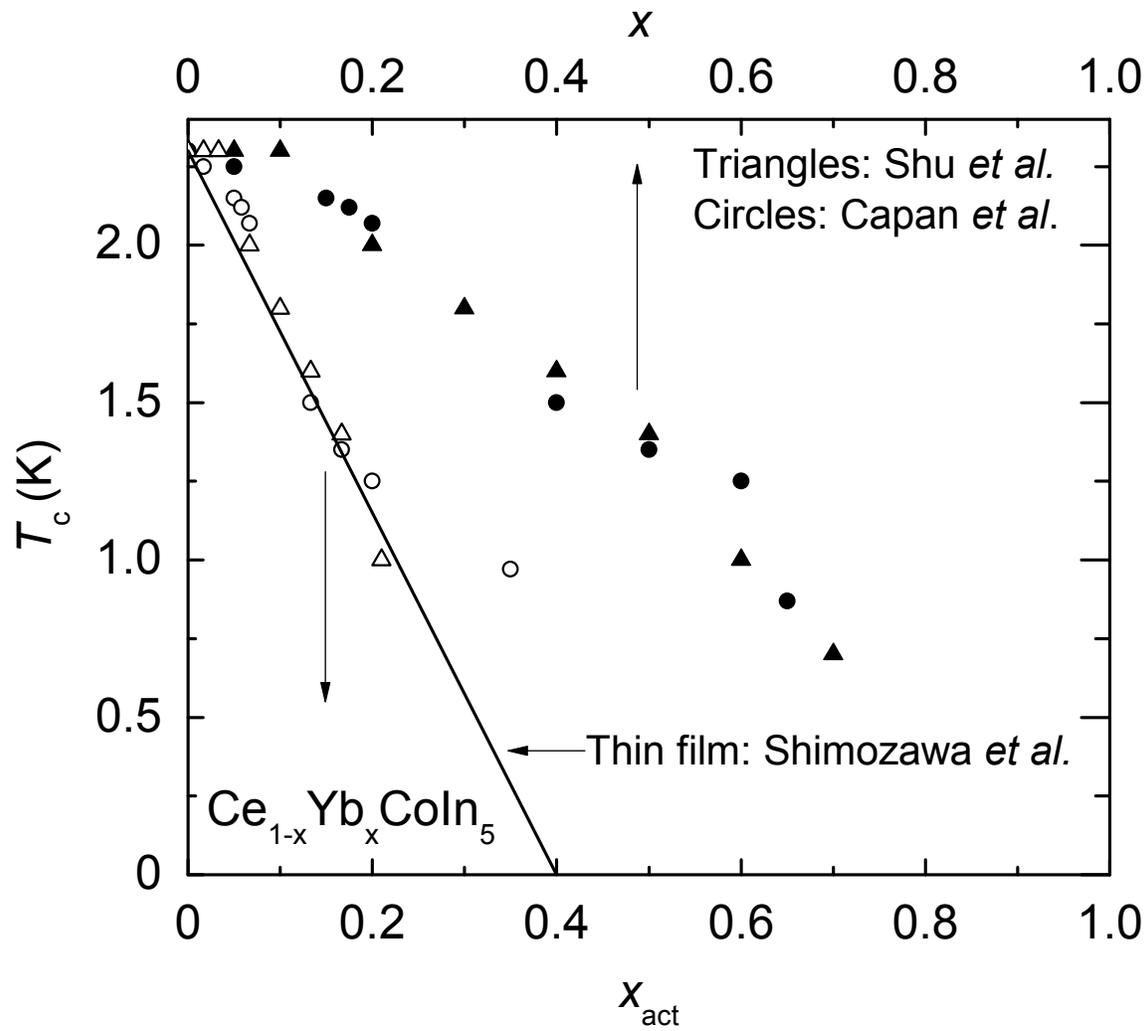

Figure 7

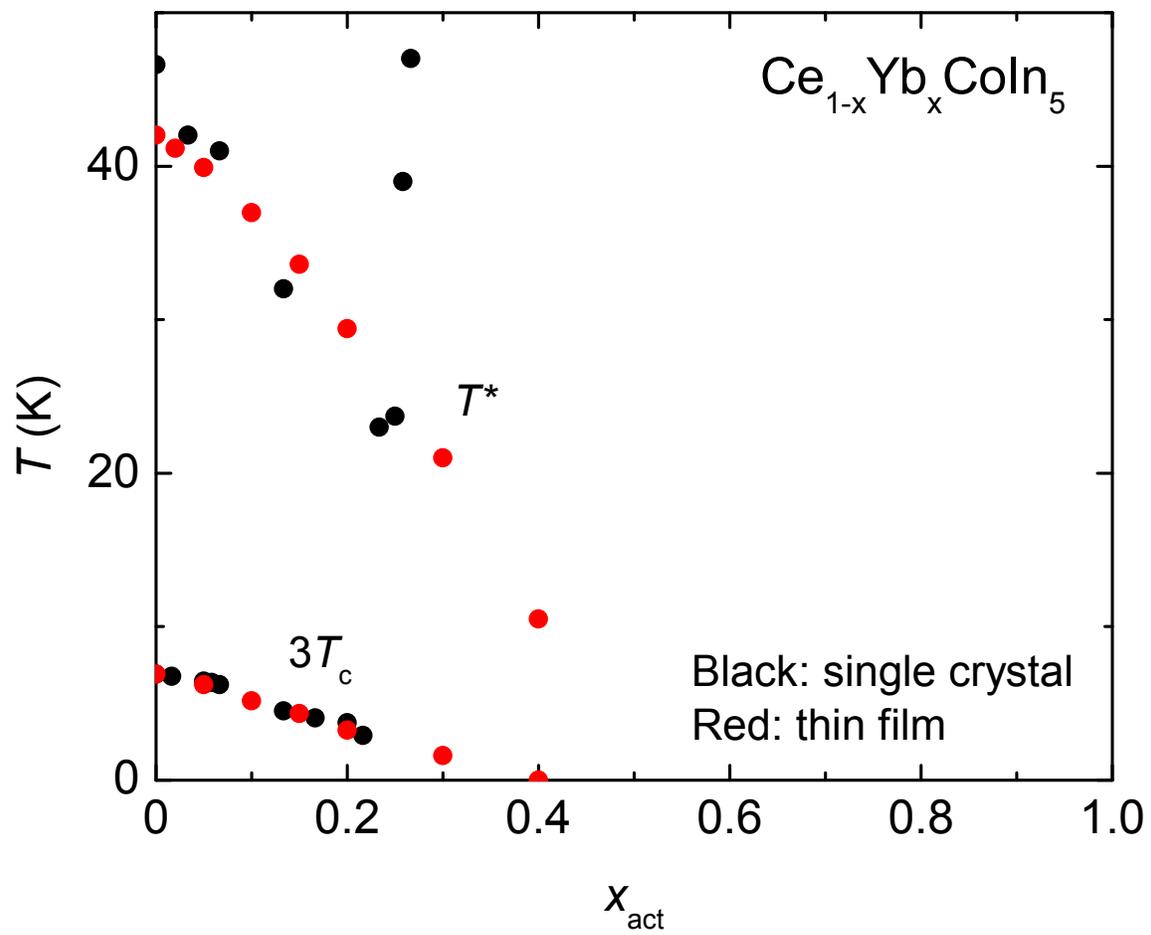